\documentclass[twocolumn,preprint]{aastex62}

\usepackage{amsmath,amstext, amssymb}
\usepackage[T1]{fontenc}
\usepackage{apjfonts} 
\usepackage{graphicx}
\usepackage{color}
\usepackage{tablefootnote}
\usepackage{subfigure}
\usepackage{natbib}
\usepackage[super]{nth}
\usepackage{multirow}

\newcommand{\Ms}{M_\star}
\newcommand{\HeI}{\hbox{{\rm He}\kern 0.1em{\sc i}}}
\newcommand{\Ha}{\hbox{{\rm H}$\alpha$}}
\newcommand{\NII}{\hbox{{\rm [N}\kern 0.1em{\sc ii}{\rm ]}}}
\newcommand{\Hb}{\hbox{{\rm H}$\beta$}}
\newcommand{\OII}{\hbox{{\rm [O}\kern 0.1em{\sc ii}{\rm ]}}}
\newcommand{\OIII}{\hbox{{\rm [O}\kern 0.1em{\sc iii}{\rm ]}}}
\newcommand{\OIIIHb}{\OIII/\Hb}

\begin{document}

\title{\large \bf CLEAR: Spatially Resolved Emission Lines and Active Galactic Nuclei at $0.6<z<1.3$}

\author[0000-0001-8534-7502]{Bren E. Backhaus}
\affiliation{Department of Physics, 196 Auditorium Road, Unit 3046, University of Connecticut, Storrs, CT 06269}

\author[0000-0002-8584-1903]{Joanna S. Bridge}
\affiliation{Department of Physics and Astronomy, 102 Natural Science Building, University of Louisville, Louisville, KY 40292}

\author[0000-0002-1410-0470]{Jonathan R. Trump}
\affiliation{Department of Physics, 196 Auditorium Road, Unit 3046, University of Connecticut, Storrs, CT 06269}

\author[0000-0001-7151-009X]{Nikko J. Cleri}
\affiliation{Department of Physics and Astronomy, Texas A\&M University, College Station, TX 77843-4242}
\affiliation{George P. and Cynthia Woods Mitchell Institute for Fundamental Physics and Astronomy, Texas A\&M University, College Station, TX 77843-4242}
\affiliation{Department of Physics, 196 Auditorium Road, Unit 3046, University of Connecticut, Storrs, CT 06269}

\author[0000-0001-7503-8482]{Casey Papovich}
\affiliation{Department of Physics and Astronomy, Texas A\&M University, College Station, TX 77843-4242}
\affiliation{George P. and Cynthia Woods Mitchell Institute for Fundamental Physics and Astronomy, Texas A\&M University, College Station, TX 77843-4242}

\author[0000-0002-6386-7299]{Raymond C. Simons}
\affil{Space Telescope Science Institute, 3700 San Martin Drive,
  Baltimore, MD, 21218 USA}

\author[0000-0003-1665-2073]{Ivelina Momcheva}
\affil{Space Telescope Science Institute, 3700 San Martin Drive,
  Baltimore, MD, 21218 USA}
\affil{Max-Planck-Institut für Astronomie, Königstuhl 17, D-69117 Heidelberg, Germany}

\author[0000-0002-4884-6756]{Benne W. Holwerda}
\affiliation{Department of Physics and Astronomy, 102 Natural Science Building, University of Louisville, Louisville, KY 40292}

\author[0000-0001-7673-2257]{Zhiyuan Ji}
\affiliation{University of Massachusetts Amherst, 710 North Pleasant Street, Amherst, MA 01003-9305, USA}

\author[0000-0003-1187-4240]{Intae Jung}
\affiliation{Astrophysics Science Division, Goddard Space Flight Center, Greenbelt, MD 20771}

\author[0000-0002-7547-3385]{Jasleen Matharu}
\affiliation{Department of Physics and Astronomy, Texas A\&M University, College Station, TX 77843-4242}
\affiliation{George P. and Cynthia Woods Mitchell Institute for Fundamental Physics and Astronomy, Texas A\&M University, College Station, TX 77843-4242}


\begin{abstract}
We investigate spatially-resolved emission-line ratios in a sample of 219 galaxies ($0.6<z<1.3$) detected using the G102 grism on the \emph{Hubble Space Telescope} Wide Field Camera 3, taken as part of the CANDELS Ly$\alpha$ Emission at Reionization (CLEAR) survey, to  measure ionization profiles and search for low-luminosity active galactic nuclei (AGN). We analyze $\OIII$ and $\Hb$ emission-line maps, enabling us to spatially resolve the $\OIII/\Hb$ emission-line ratio across the galaxies in the sample. We compare the \OIIIHb\ ratio in galaxy centers and outer annular regions to measure ionization gradients and investigate the potential of sources with nuclear ionization to host AGN.  
We investigate some of the individual galaxies that are candidates to host strong nuclear ionization and find that they often have low stellar mass and are undetected in X-rays, as expected for low-luminosity AGN in low-mass galaxies. We do not find evidence for a significant population of off-nuclear AGN or other clumps of off-nuclear ionization. We model the observed distribution of \OIIIHb\ gradients and find that most galaxies are consistent with small or zero gradients, but 6-16\% of galaxies in the sample are likely to host nuclear \OIIIHb\ that is $\sim$0.5~dex higher than in their outer regions. This study is limited by large uncertainties in most of the measured \OIIIHb\ spatial profiles, therefore deeper data, e.g, from deeper \textit{HST}/WFC3 programs or from \textit{JWST}/NIRISS, are needed to more reliably measure the spatially resolved emission-line conditions of individual high-redshift galaxies.
\\

\end{abstract}


\section{Introduction}

The rest-frame optical spectra of galaxies can be used to infer a wide variety of physical properties via their nebular and recombination lines to determine, for example, star formation rate \citep[SFR; e.g.,][]{kennicutt1998, kennicutt2012}, metallicity \citep[e.g.,][]{pagel1981, zahid2013, maiolino2019}, interstellar medium (ISM) density \citep{kewley2019}, temperature \cite[e.g.,][]{peimbert2017}, and dust attenuation \citep[e.g.,][]{cardelli1989, calzetti2001}. Additionally, emission-line ratios have been used to identify active galactic nuclei (AGN), such as the $\NII/\Ha$ and $\OIII/\Hb$ ratios used in ``BPT''\footnote{Baldwin, Phillips, and Terlevich} diagram \citep{baldwin1981, veilleux1987, kewley2001, kauffmann2003}, Mass-Excitation (MEx) diagram \citep{juneau2011}, and the OIII/Hb vs NeIII/OII (OHNO) ``OHNO'' diagram \citep{back22}. Narrow-line AGN selection methods are particularly important because they are complementary to multi-wavelength AGN selections: different methods have different biases as a function of host galaxy properties and/or AGN Eddington ratio \citep{hickox2009, buchner2015, aird2012,jones2016, cann2019, Trump13blAGN, Ji2022, Lambrides2020, trump2015}.

Determining the physical conditions of galaxies around the peak of cosmic star formation ($z\sim2$) is complicated. At these redshifts, optical diagnostic lines such as \OIII$\lambda$5007+$\lambda$4959 and H$\alpha$ are redshifted into the infrared, where atmospheric opacity and high background make ground-based observations challenging. Additionally, interpreting BPT diagrams at higher redshifts is complicated by higher star formation rates \citep{madau2014} and harder ionizing radiation \citep{kewley2015, steidel2014}, making it difficult to distinguish ionization from an AGN and from high-redshift star formation processes \citep{moran2002, trump2015, coil2015}. This is especially true in low-mass galaxies where X-ray detection of AGN is unlikely \citep{xue2010, aird2012}.

Spatially resolved lines can be used to gather further information on galaxy evolution and formation. Spatially resolving $\Ha$ gives the profile of star formation within a galaxy. This gives us information whether a galaxy is formed inside-out or outside-in by comparing to the galaxy's stellar continuum \citep{nelson2012, nelson2016a, nelson2021, matharu2022}. Spatially resolving the Balmer decrement ($\Ha$/$\Hb$) measures the profile of dust attenuation within a galaxy. In low-mass galaxies at the redshifts $z\sim1.4$ the dust attenuation gradients have been measured to be relatively flat on average \citep{nelson2016b}. As mass increases, dust attenuation increases toward the center of the galaxy \citep{nelson2016b}.

Spatially resolving the \OIIIHb\ line ratio has shown promise for decomposing nuclear AGN and extended star formation activity. \cite{trump2011,wright2010} examined stacked WFC3/G141 grism spectra and found that elevated central \OIIIHb\ ratios may indicate the presence of obscured or dim AGN. Using simulations of 2-orbit WFC3/G141 grism observations, \cite{bridge2015} showed that it is possible to detect low-luminosity AGN in individual galaxies, particularly in low-mass galaxies, by spatially resolving the \OIIIHb\ line ratio in the inner and outer regions of the galaxy using the G141 spectrum. That study made use of the Mass-Excitation (MEx) diagnostic diagram \citep{juneau2011, juneau2014} in order to distinguish between AGN and star formation activity. The MEx diagram replaces the [NII]/H$\alpha$ axis of the traditional BPT diagram with stellar mass. The mass-metallicity relation has shown that there is a correlation between mass and [NII]/H$\alpha$ \citep[e.g.,][]{tremonti2004}, facilitating the use of this diagnostic even when the [NII] and H$\alpha$ lines are not available.



In order to address the challenges of understanding high-redshift emission-line galaxies, the Wide Field Camera 3 \citep[WFC3;][]{kimble2008} G102 ($0.8<\lambda<1.15~\mu$m, $R\sim210$) and G141 ($1.1<\lambda<1.7~\mu$m, $R\sim130$) slitless grisms on the \emph{Hubble Space Telescope} (\emph{HST}) have been used to conduct large surveys of thousands of galaxies in the near infrared regime. These surveys have expanded our understanding of physical conditions of galaxies since cosmic noon [e.g., the 3D-HST \citep{momcheva2016} and Faint Infrared Grism Survey (FIGS; \citealt{pirzkal2017}) surveys]

In this work we investigate the spatially resolved \OIIIHb\ emission-line profiles of 219 galaxies at redshift $z\sim0.9$ observed by the CLEAR survey. We use the spatially resolved emission-line ratios to infer the ionization profiles of the galaxies and particularly search for nuclear ionization in low-mass galaxies in order to identify
low-luminosity AGN that are typically missed by X-ray detection.
We also compare galaxy ionization profiles to other galaxy properties such as stellar mass, star formation rate, effective radius, and redshift.

In Section 2, we describe the data observations and reduction. In Section 3, we explain our sample selection, and in Section 4 we discuss our method for spatially resolving the \OIIIHb\ line ratio. Section 5 describes how the  spatially resolved \OIIIHb\ emission-line relates to different galaxy properties. We discuss the spatially resolved \OIIIHb\'s ability to detect low luminosity AGN and regions of higher ionization in Section 6. We present our conclusions in Section 7. Throughout this paper, we assume a $\Lambda$CDM cosmology with $H_0 = 70$ km s$^{-1}$ Mpc$^{-1}$, $\Omega_m = 0.3$, and $\Omega_\Lambda = 0.7$ \citep{plan15} .

\section{Observations and Data Reduction}

The data used in this work is comprised of \emph{HST}/WFC3 G102 grism $(0.8 < \lambda < 1.15 ~\mu$m) observations from various surveys, the majority of which was taken as part of the CANDELS Ly$\alpha$ Emission at Reionization (CLEAR; \citealp{estrada19}) G102 grism survey. The G102 observations come from programs GO-14227 (PI: C. Papovich), GO-13420 (PI: G. Barro), and GO/DD-11359 (‘ERS’, PI: R. O’Connell). The \emph{HST}/WFC3 also has G141 observations which come from programs GO-11600 (‘AGHAST’; PI: B. Weiner), GO-12461 (‘SN Colfax’, PI: A. Reiss), GO-13871 (PI: P. Oesch), GO/DD-11359 (‘ERS’, PI: R. O’Connell), GO12099 (‘George, Primo’, PI: A. Reiss), and GO-12177 (‘3D-HST’, PI: van Dokkum). CLEAR includes six pointings in the CANDELS \citep{koekemoer2011} Great Observatories Origin Deep survey (GOODS)-North (GN) field with 10 orbit depth and six pointings in the GOODS-South (GS) field with 12 orbit depth. These datasets provide low-resolution grism spectroscopy over observed-frame $0.8<\lambda<1.65$~\micron\ for every source in the field of view,  with $R \sim 210$ in G102. The CLEAR pointings overlap with the 3D-HST survey \citep{momcheva2016}, which gives G141 slitless grism spectra covering $1.1-1.65$~\micron\ with a 2-orbit depth which were used in the data reduction to aid in redshift determination. In contrast to the low spectral resolution, the two-dimensional spectra have high \textit{spatial} resolution of $0\farcs06$ per pixel.

The CLEAR survey is augmented by G102 observations from the Faint Infrared Grism Survey (FIGS; HST-GO 13776, PI Malhotra). The FIGS campaign is comprised of four pointings of 40-orbit observations in the GOODS fields \citep[see][]{tilvi2016}, one of which overlaps with a CLEAR pointing in GOODS-S. This pointing was therefore also ingested into the CLEAR data. The FIGS data were reduced and processed in the same fashion as the CLEAR data (Section 2.2).

The CLEAR spectra have been used to study the metallicities, ages, and formation histories of massive high-redshift galaxies (\citealt{estrada19,Estrada2020, simons21}), and to appraise Paschen-$\beta$ as a star-formation rate indicator in low-redshift galaxies \citep{Cleri22}. \cite{matharu2022} used spatially resolved H-alpha emission line maps of star-forming galaxies to study the evolution of gradients in galaxy assembly. \cite{jung2021} used CLEAR to study the evolution of the strength of Lyman Alpha emission at 6.5<z<8.2. \cite{back22} and \cite{Papovich2022} used CLEAR spectra to study the physical conditions of galaxies and their (spatially integrated) gas conditions.

\subsection{CLEAR Observing Strategy}
The CLEAR observations were taken over the course of two years (November 2015 to February 2017). Direct imaging observations were done using the WFC3 F105W filter. Each field was observed with three different orients, four orbits for each orient in GS and GN has two four orbit orients with a single two orbit orient, 
separated by $\sim10-20$ degrees in roll angle to facilitate the separation of overlapping spectra.

The Earth's atmosphere produces a time-variable background caused by $\HeI$ emission at $10830$~\AA\ that affects both the F105W filter and the G102 grism. The observations were scheduled to reduce this background by placing the G102 grism observations during times of low background and the F105W observations when the background is predicted to be higher in order to minimize the background in the spectra (see \citealt{brammer2014, tilvi2016, lotz2017} for further information.)\\

\subsection{Spectroscopic Data Reduction}

The G102 grism data from all the surveys were reduced together using the grism redshift \& line analysis software \texttt{grizli} \citep{grizli}. The complete reductions are described in detail in \citep{simons21}, but we describe the process briefly here.
CLEAR F105W reference images and dispersed G102 grism were aligned to the same world coordinate system (WCS) as 3DHST photometric catalog \citep{skelton2014}.
Cosmic ray cleaning, flat-fielding, and sky subtraction (using WFC3/IR master sky images, \citealp{brammer2015}) was then performed.

Contamination from overlapping spectra is a known issue in grism spectroscopy. It is more significant for faint sources contaminated by bright ones because subtracting bright sources from faint spectra introduces errors and contributes residual flux to the spectra of faint sources.
The continuum model of these overlapping sources are subtracted from object spectra. The contamination of the 2D spectra was modeled in two steps. First, a flat continuum model was used for objects down to $m_{\textrm{F105W}} < 25$. Subsequently, a polynomial continuum was used for objects with $m_{\textrm{F105W}} < 24$. All objects with $m_{\textrm{F105W}} < 25$ were then extracted from the CLEAR imaging, with 2D spectral extractions for each grism observation, for about 6000 extracted objects. 

\begin{figure*}[t]
\centering
\epsscale{0.9}
\plotone{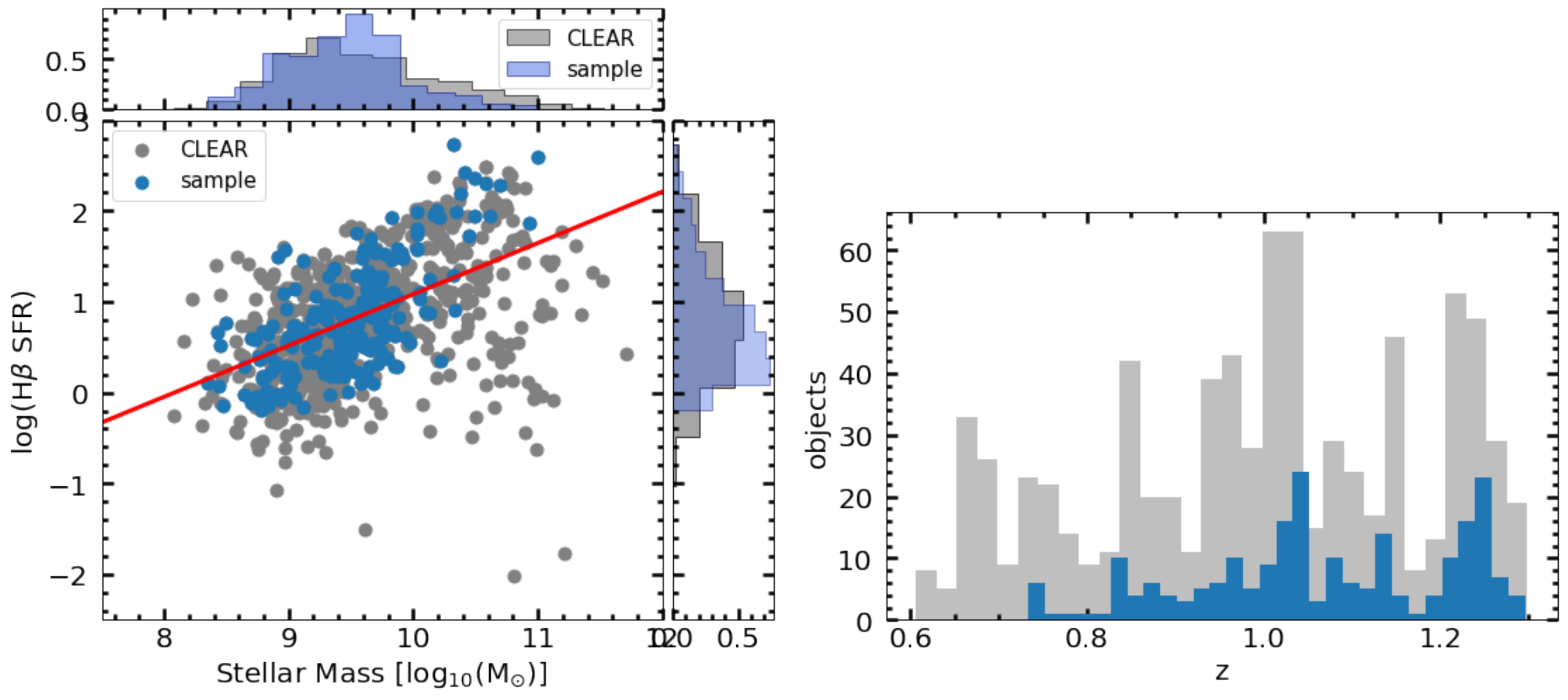}
\caption{Left: The distribution of stellar mass and $\Hb$ SFR from our selected sample of CLEAR galaxies between $0.6<z<1.3$, color-coded by redshift, compared to the complete CLEAR data set between $0.6<z<1.3$ (grey points). Our sample contains about 31\% of the CLEAR galaxies at this redshift. The red line is the star formation mass sequence from \cite{whitaker2012} shown in Equation \ref{Eq:SFMS}. Right: The distribution of redshifts from our selected sample of CLEAR galaxies between $0.6<z<1.3$ (blue points) compared to the complete CLEAR data set between $0.6<z<1.3$ (grey points).
\label{fig:sample}}
\end{figure*} 

Fits to the spectra were carried out with \texttt{grizli} using the CLEAR F105W images as well as the photometry available from 3D-HST \citep{skelton2014}. In order to determine the redshift of each object, Flexible Stellar Population Synthesis (FSPS) stellar population models \citep{conroy2010} were used with fixed emission line sets and ratios. Line fluxes were then determined using the best-fit redshift.

The \texttt{grizli} software creates spatially resolved emission-line maps by fitting a model to the 2D spectrum of the object of interest, and subtracting the continuum. The line maps for each emission line present in the spectrum are then created by drizzling the best-fit galaxy model at the wavelength of each emission line in the spectrum. This results in a 2D map with a pixel scale of $0.1 \times 0.1$~arcsec of the flux that is in excess of the continuum at each line wavelength.

\section{Sample selection}

The final CLEAR catalog (Simons et al. in prep) contains the 1D and 2D spectra, total (integrated) line fluxes, and emission-line maps of more than 6000 objects. We implemented a cut in the grism redshift fits to require $(z_{97}-z_{02})/2 < 0.005$ to ensure robust line identification, where $z_{97}$ and $z_{02}$ are the \nth{97} and \nth{2} percentiles of the redshift probability distribution produced by \texttt{grizli} for each object. 

We focus on the \OIIIHb\ ratio (with air wavelengths of 5007 and 4861~\AA, respectively) that is ideal for detection with the G102 grism. This ratio has been identified as a tracer of AGN activity \citep[e.g., ][]{baldwin1981, juneau2014, bridge2015} because a high \OIIIHb\ ratio indicates the presence of the harder ionizing radiation necessary to produce \OIII\ 
oxygen \citep{kewley2015}. Additionally, the fact that the $\OIII$ doublet and $\Hb$ are close in wavelength makes their ratio nearly independent of extinction. Requiring multiple line pairs (such as [NeIII]/[OII] or [SII]/$H\alpha$) would significantly reduce the redshift range and sample size.

\begin{figure*}
\centering
\begin{subfigure}
    \centering
	{\includegraphics[width=0.7\textwidth]{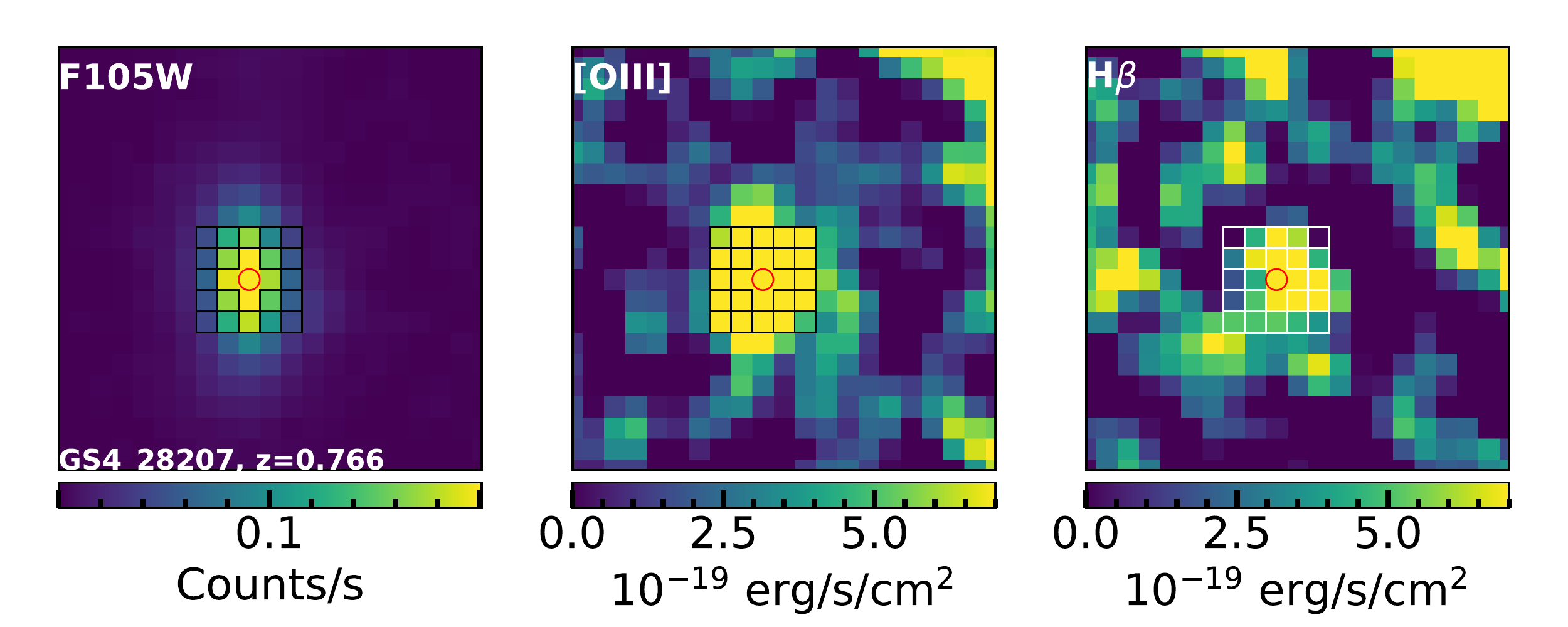}}
\end{subfigure}
\begin{subfigure}
	\centering
	{\includegraphics[width=0.71\textwidth]{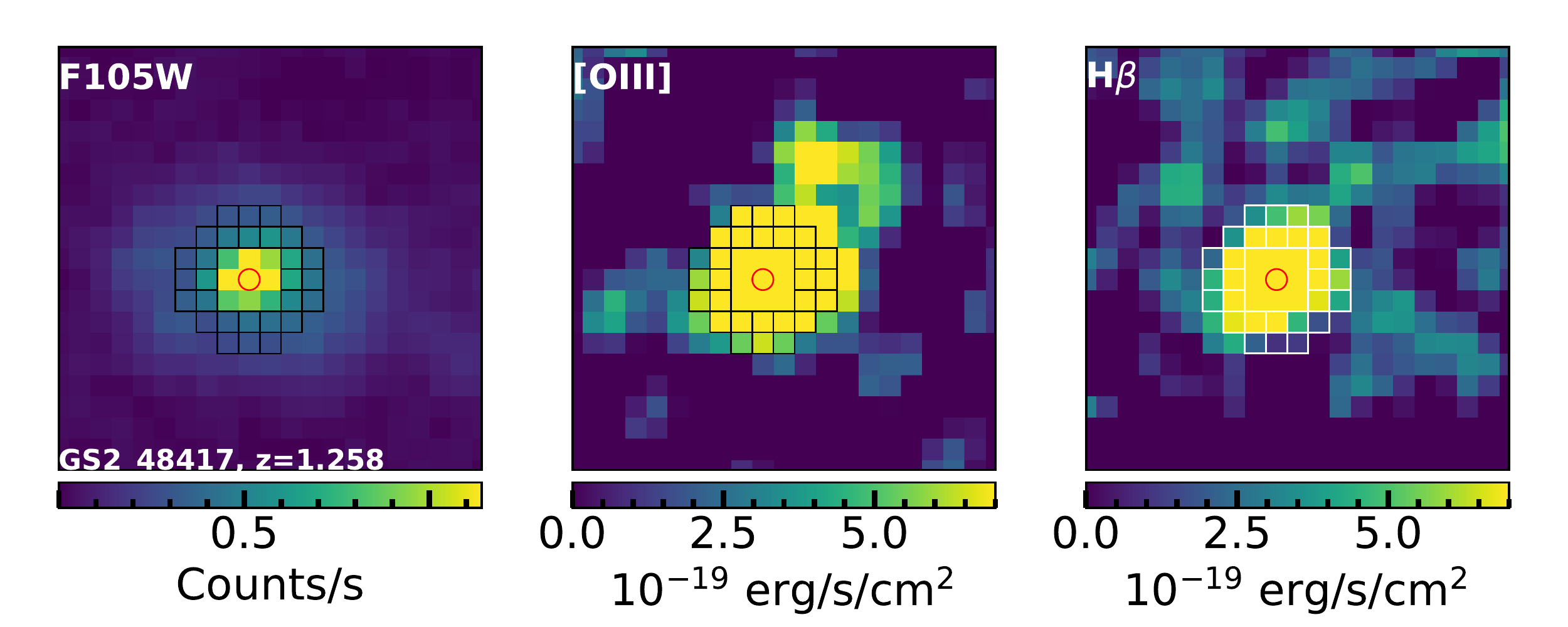}}
\end{subfigure}
\caption{The upper and lower panels show examples of a large and small galaxy and their outer-aperture regions. \emph{Top:} A small galaxies with $R_{50} < 2\farcs5$ with an outer-aperture region of 20 pixels shown as boxes. \emph{Bottom:} A large galaxy with  $R_{50} > 2\farcs5$ with an outer-aperture region of 28 pixels shown as boxes. \emph{Left}: The F105W direct images. \emph{Center}: The $\OIII$ emission-line maps. \emph{Right}: The $H\beta$ emission-line maps. The pixel scale is $0\farcs1$ in all panels, corresponding to $\sim$1~kpc at the redshift of these galaxies. The center of these galaxies is the F105W centroid.}
\label{ex_moveup}
\end{figure*}

We then use the integrated $\OIII$ and $\Hb$ line fluxes (total fluxes integrated across a whole galaxy) to impose a signal-to-noise cut on the \OIIIHb\ emission line ratio of S/N$_{\textrm{[O~III]/H}\beta} > 3/\sqrt{2}$, the low end of which is equivalent to both lines being measured at exactly $3\sigma$. Implementing a signal-to-noise cut on the ratio ensures that either $\OIII$ \emph{or} $\Hb$ are measured by at least 3$\sigma$, but allows for the inclusion of objects where one line may be less well-measured than the other. The resulting sample is almost 30\% larger than that generated by a traditional ${\rm S/N} > 3$ cut on the $\OIII$ line flux measurement.

Finally, we examine both the 1D spectra and 2D emission-line maps by eye to ensure a robust sample of galaxies with well-determined \OIIIHb\ ratios. In some cases, errors in the continuum model leave unphysical artifacts in the emission-line maps. We remove these objects from the sample to include only those with well-fit continua. This removes $\sim 28\%$ of our sample leaving our final sample to contain 219 emission-line objects. The $\Hb$ SFR and stellar mass distribution of our sample compared to the entire CLEAR sample at $0.6<z<1.3$ is shown in Figure \ref{fig:sample}.
This SFR is corrected for dust attenuation using the $\Ha/\Hb$ ratio.

The star formation mass sequence line is from \cite{whitaker2012} using the sample's average redshift of $z=1.0$:

\begin{equation} \label{Eq:SFMS}
\begin{split}
  \log({\rm SFR})[M_\odot/\mathrm{yr}] = \alpha(z)(\log(M_{*})-10.5)+\beta(z)  \\
  \alpha(z)=(0.7-0.13z) \\
  \beta(z)=0.38+1.14z-0.19z^2
 \end{split}
\end{equation}

\subsection{Stellar Masses}

Following the method of \citet{estrada19}, the stellar masses were determined using \texttt{eazy-py} \citep{eazy-py}, a spectral energy distribution (SED) fitting package based on the photometric redshift fitting code \texttt{EAZY} \citep{brammer2008}. The algorithm uses 12 FSPS templates, each with their own unique star formation history. The templates all use a Chabrier initial mass function \citep{chabrier2003}.

\subsection{X-ray Counterparts}
We matched objects in our sample with known X-ray-emitting counterparts in the Chandra Deep Fields using the the 2-Ms CDF-N 
\citep{xue2016} and 7-Ms CDF-S \citep{luo2017} point-source catalogs.
Any galaxy from the CLEAR sample that corresponds to an optical counterpart in the X-ray source catalogs within 0\farcs5 is considered an X-ray object. Two counterparts were classified as X-ray galaxies (i.e., X-ray emission consistent with X-ray binaries and other processes associated with star formation) and 11 objects were classified as X-ray AGN. These classifications were made based on the strength and hardness of the galaxies' X-ray flux (see \citealt{xue2016} and \citealt{luo2017} for details).

\begin{deluxetable*}{|l|r|r|r|r|r|r|r|}[t]
\tablecaption{Sample \label{tab:sample}}
\tablenum{1}
\tablecolumns{8}
\tablewidth{0pt}
\tablehead{\colhead{ID} &  \colhead{Global Ratio} &   \colhead{Inner Ratio} &  \colhead{Outer Ratio} & \colhead{$\log(\Ms/M_\odot)$} &    \colhead{$z$} &  \colhead{log(SFR)} &   \colhead{$R_{50}$}}
\startdata
$GN1\_36795$ &  $4.42 \pm 1.14$ &    $1.74 \pm 1.45$ &     $1.99 \pm 0.93$ &  9.87 & 1.22 &    1.58 & 0.39 \\
\hline
$GN1\_38027$ &  $2.50 \pm 0.69$ &    $1.00 \pm 0.50$ &     $2.06 \pm 2.14$ &  9.39 & 0.98 &    0.18 & 0.27 \\
\hline
$GN1\_37494$ &  $2.31 \pm 0.55$ &    $0.69 \pm 0.51$ &     $1.08 \pm 0.29$ &  9.54 & 1.02 &    0.59 & 0.52 \\
\hline
$GN1\_37031$ &  $0.84 \pm 0.15$ &    $0.32 \pm 0.24$ &     $0.28 \pm 0.31$ &  9.76 & 1.06 &    1.08 & 0.58 \\
\hline
$GN1\_37567$ &  $2.88 \pm 0.40$ &    $3.05 \pm 1.21$ &     $7.80 \pm 7.55$ &  9.63 & 1.18 &    0.96 & 0.29 \\
\hline
$GN1\_37812$ &  $2.08 \pm 0.84$ &    $0.70 \pm 0.74$ &     $0.72 \pm 0.50$ &  9.72 & 0.78 &    1.20 & 0.55 \\
\hline
$GN1\_37691$ &  $4.46 \pm 1.15$ &    $0.84 \pm 0.59$ &     $3.12 \pm 1.93$ &  9.49 & 1.15 &    0.47 & 0.40 \\
\hline
$GN1\_37700$ &  $1.65 \pm 0.40$ &    $1.37 \pm 0.47$ &     $1.09 \pm 0.67$ &  9.88 & 1.04 &    0.84 & 0.15 \\
\hline
$GN1\_37053$ &  $1.44 \pm 0.23$ &    $2.74 \pm 2.89$ &     $1.17 \pm 0.42$ &  8.87 & 1.04 &    0.74 & 0.32 \\
\hline
$GN1\_37750$ & $12.21 \pm 4.00$ &    $6.03 \pm 1.56$ &     $3.55 \pm 1.56$ &  8.43 & 1.23 &    0.66 & 0.05 \\
\enddata
\caption{The \OIIIHb\ ratios for the galaxies in the sample, measured from the global profile (e.g., integrated over the whole galaxy), in the inner central pixel, and in the outer aperture.
Other columns indicate the stellar mass, redshift, SFR and effective radius of the galaxies.
Only a portion of this table is shown here to demonstrate its form and content. A machine-readable version of the full table is available.}
\end{deluxetable*}

\begin{deluxetable*}{|l|r|r|r|r|r|r|r|}[t]
\tablecaption{Binned Data \label{tab:mbin}}
\tablenum{2}
\tablecolumns{8}
\tablewidth{0pt}
\tablehead{\colhead{$\log(\Ms/M_\odot)$} & \colhead{$\log(\Ms/M_\odot)_{Med}$} &   \colhead{$\log(SFR)_{Med}$} &  \colhead{$N_{gal}$} &  \colhead{Global Ratio} &  \colhead{Inner Ratio} &  \colhead{Outer Ratio} &      \colhead{Gradient} }
\startdata
\multirow{2}{4em}{8.34-8.99} &8.79 &   -0.01  &     21 &    0.60$\pm$0.01 &      0.51 $\pm$0.02&       0.33$\pm$0.05&  0.18$\pm$0.06  \\ 
    &  8.86 &  0.59 &     22 &       0.67$\pm$0.01 &      0.64 $\pm$0.04&       0.39$\pm$0.05 &  0.25 $\pm$0.07 \\ 
\hline
\multirow{2}{4em}{9.00-9.34} & 9.18 &   0.28 &     21 &      0.64$\pm$0.02  &     0.56$\pm$0.05 &       0.46 $\pm$0.07& 0.09 $\pm$0.08 \\
    & 9.20 &   0.77&     22 &   0.62 $\pm$0.01&      0.40 $\pm$0.05&       0.48 $\pm$0.07& -0.08 $\pm$0.08  \\
\hline
\multirow{2}{4em}{9.36-9.58} &   9.42 &  0.33  &     22 &      00.54$\pm$0.02 &      0.26$\pm$0.03 &       0.36$\pm$0.04 & -0.10$\pm$0.05 \\
    &   9.49 &   0.86  &     23 &      0.55$\pm$0.01 &      0.28 $\pm$0.05&       0.32$\pm$0.05 & -0.05$\pm$0.07\\
\hline
\multirow{2}{4em}{9.59-9.79} &   9.69 &   0.59  &     22 &         0.34 $\pm$0.02&      0.15$\pm$0.07 &       0.20$\pm$0.06 & -0.04$\pm$0.09  \\
     &   9.66 &   1.21 &     22 &      0.49 $\pm$0.01&      0.33$\pm$0.02 &       0.29$\pm$0.04 &  0.03$\pm$0.05 \\
\hline
\multirow{2}{4em}{9.82-10.99} &   9.96 &  0.94  &     22 &      0.29$\pm$0.02 &     -0.00 $\pm$0.03&       0.19 $\pm$0.04& -0.19 $\pm$0.05 \\
    &  10.33 &   1.95  &     22 &      0.31$\pm$0.02 &  0.16 $\pm$0.03&       0.18 $\pm$0.04  & -0.02$\pm$0.05 \\
\enddata
\caption{Median \OIIIHb\ emission-line ratios for galaxies in bins of stellar mass and SFR. We first break our data into five stellar mass bins, and we use the median SFR of the mass bin to create two sub-bins representing high and low SFRs. The uncertainty of the median is given for the ratio and gradient measurements. The low (high) SFR bins are depicted as stars (hexagons) in Figures \ref{MEx}, \ref{fig:invout}, and \ref{fig:movers}. }
\end{deluxetable*}

\section{Spatially Resolved Line Ratios}

\begin{figure}[t]
\centering
\epsscale{1.1}
\plotone{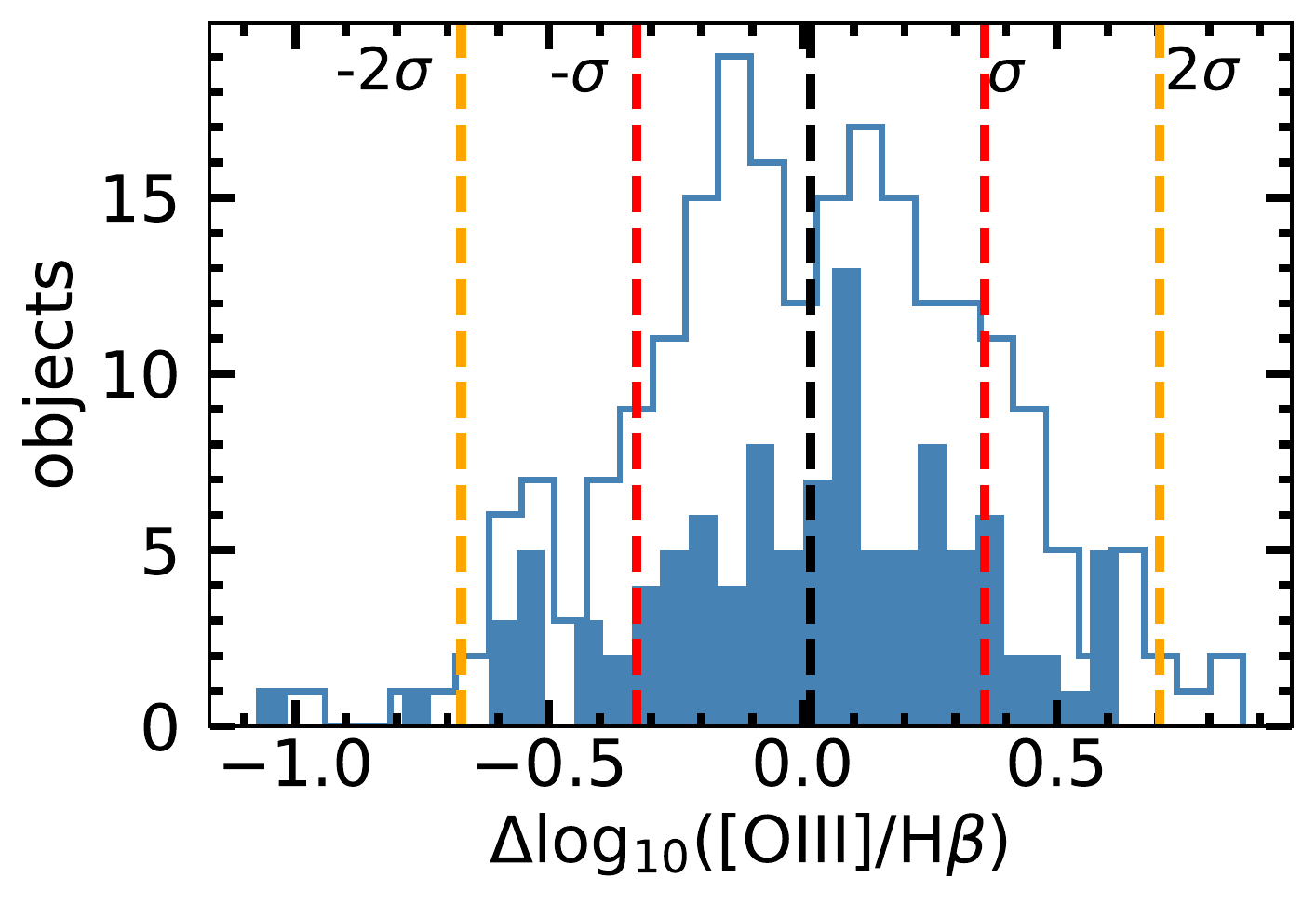}
\caption{Distribution of the inner minus the outer \OIIIHb\ ratio. The solid histogram represents the distribution of galaxies that have well-measured \OIII\ and \Hb\ fluxes in both their central pixel and outer aperture. The open histogram represents the distribution of the entire sample including galaxies that have an upper limit in one or more of the central or outer \OIII\ or \Hb\ fluxes. The median gradient is 0.0147 and standard deviation is 0.3428, and most galaxies have a gradient that is consistent with zero.}
\label{fig:grad}
\end{figure} 

\begin{figure}[t]
\centering
\epsscale{1.1}
\plotone{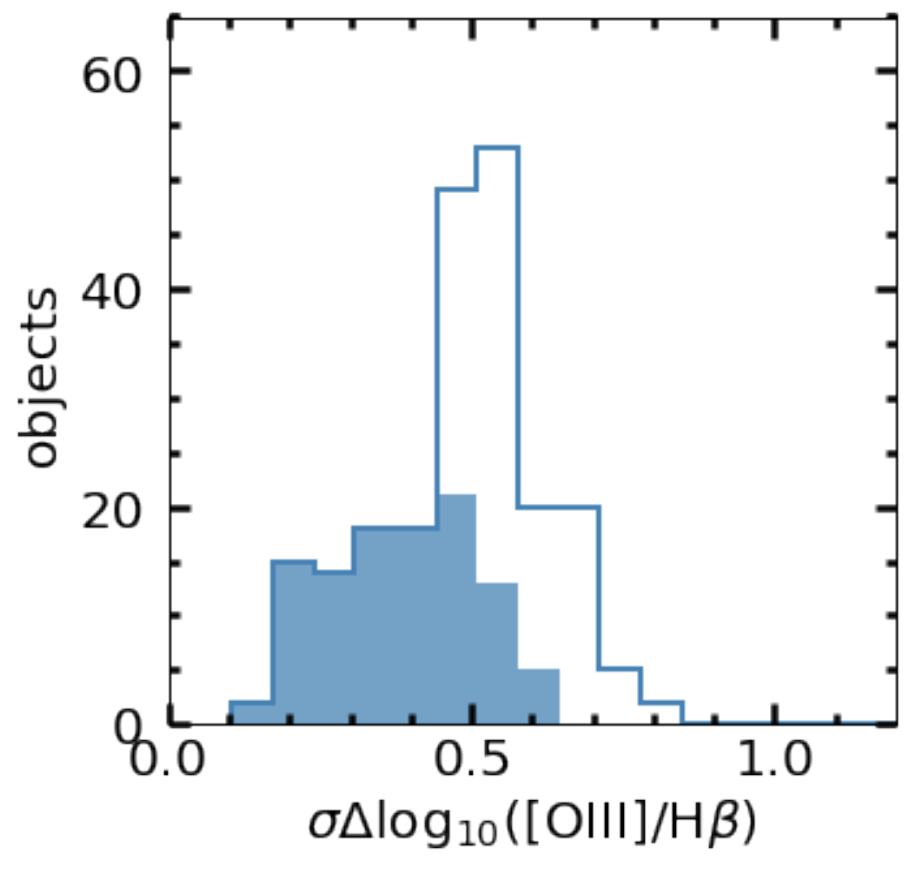}
\caption{Distribution of the uncertainty in the inner minus the outer \OIIIHb\ ratio. The solid histogram represents the distribution of galaxies that have well-measured \OIII\ and \Hb\ fluxes in both their central pixel and outer aperture. The open histogram represents the distribution of the entire sample including galaxies using a limit of \OIII\ or \Hb. As expected,galaxies with well-measured fluxes have smaller uncertainties. In general most of the line-ratios have large uncertainties compared to the measurement.} 
\label{fig:grad_uncer}
\end{figure} 

We use the 2D line maps to measure the spatially-resolved emission lines from each galaxy. We are specifically interested in the \OIIIHb\ line ratio, and we use the line maps (examples of which are shown in Figure~\ref{ex_moveup}) to measure differences in the ratios across the spatial profile of each galaxy. We begin by defining the ``center'' of the galaxies, i.e., where we expect to find a central AGN, to be the middle of the pixel of the photometric centroid of the F105W image. The pixel scale of both the images and the 2D grism spectra is 0\farcs1/pixel, which for the redshift range of this sample corresponds to approximately 1~kpc/pixel. This resolution is sufficient such that the ionization from a central AGN can be spatially distinguished from the more extended emission associated with distributed star formation.

In order to characterize the emission lines from the extended star-forming region of the galaxy, we define an outer aperture region outside of the central pixel. The number of pixels chosen depends on the size of the galaxy as determined from the half-light radii ($R_{50}$) from the F125W CANDELS images \citep{vanderwel2014}. For galaxies with $R_{50} >$ 2\farcs5, we selected a set of 28 pixels symmetrically around the center pixel from the extended regions of the galaxy. Smaller galaxies use a similar shape but slightly smaller and with only 20 pixels. In both cases we leave a gap of at least one pixel between the central pixel and the extended region
in order to minimize blending 
between the regions of emission and avoid correlated fluxes from neighboring pixels in the drizzled WFC3 data. The geometry of the extended region pixel selection from larger and compact objects is demonstrated in Figure~\ref{ex_moveup}.

The extended \OIIIHb\ ratio is calculated as the median \OIIIHb\ ratios in the outer aperture. We choose the median as a statistically robust measure of the extended emission-line flux, since many galaxies contain pixels with lower signal-to-noise measurements and/or outlier flux caused by contamination. Use of the median allows the outlier pixels to be ignored in the aperture measurement for the galaxy. We also calculated the outer-aperture emission-line flux using the mean, weighted mean, weighted mean with outlier rejection, and weighted median. These other methods generally resulted in similar flux values but tended to be more affected by outliers or had larger uncertainties than using the simple median. 

Our samples integrated, inner aperture, and outer aperture ratio measurements are shown in Table \ref{tab:sample} along with the galaxies stellar mass, redshift, SFR, and effective radius.

\subsection{Binned Line Ratio Profiles}

Of our sample of 219 galaxies, 113 of them have at least one limit in a flux measurement from the inner or outer regions, usually from the $\Hb$ emission line, resulting in a limit in the \OIIIHb gradient. 
We mitigate the large uncertainties and limits of individual galaxies by putting our sample, including galaxies with limits in their emission lines, into five bins based on their stellar mass, with each bin containing about 43-45 galaxies. We then split each mass bin into two sub-bins of high and low $\Hb$ star formation rates using the median $\Hb$ SFR of each bin, giving us a total of 10 bins with $22 \pm 1$ galaxies in each. 

We then calculate the median stellar mass and SFR of each bin. The $\OIIIHb$ gradient for each bin is created by taking the median emission-line fluxes from both apertures to create the ratios. 
The median mass, SFR, emission-line ratios, and gradient for each bin is shown in Table \ref{tab:mbin}.

\section{Results}

We define the ``$\OIIIHb$ gradient'' as the difference between the $\log(\OIIIHb)$ ratio in the inner pixel minus the ratio in the outer region. Figures \ref{fig:grad} and \ref{fig:grad_uncer} show the distributions of the $\OIIIHb$ gradient and $\OIIIHb$ gradient uncertainty, respectively. In these diagrams the solid blue histogram is the distribution of galaxies with well-measured emission-line fluxes in both the inner and outer apertures. The open blue histogram is the distribution of the entire sample, including galaxies with limits in at least one emission-line flux. 

Figure \ref{fig:grad} shows most of the objects have measured \OIIIHb\ gradients close to zero, although there is large scatter. 
Much of this scatter is likely due to the large gradient uncertainties seen in Figure \ref{fig:grad_uncer}:
only 29 galaxies, $13\%$ of the sample, have a gradient uncertainty less than 0.3~dex.
The median uncertainty for the entire sample is 0.5~dex (open histogram), with a median gradient uncertainty of 0.38~dex for the well-measured sample of galaxies with no limits in inner/outer emission-line fluxes (solid blue histogram).

\subsection{General Emission-Line Ratio Properties}

\begin{figure*}
    \centering
    \includegraphics[width=0.9\textwidth]{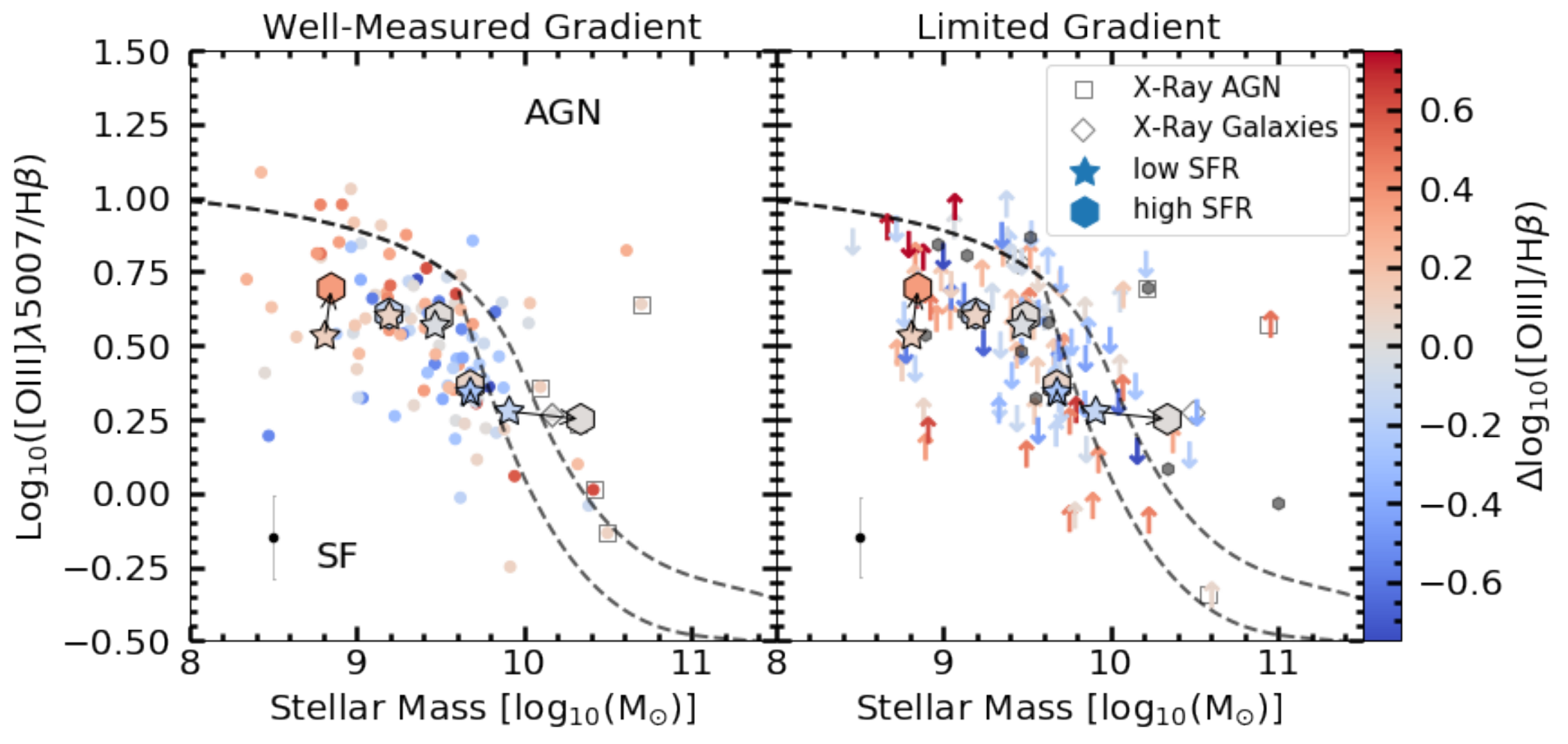}
    \caption{The MEx diagram of the galaxies in our sample. The integrated \OIIIHb\ ratios are given on the vertical axis, and the points are shaded by the inner minus the outer \OIIIHb\ ratio. Left: Galaxies with well-measured gradient in both inner and outer apertures. Right: Galaxies with limits in  an emission-line flux in the inner and/or outer regions.
    The arrows indicate an upper or lower limit in the \emph{ratio gradient}. The black circles are galaxies that have limits in both the inner and outer regions and are statistically unconstrained. The stars and hexagons in both panels are the ratios of the galaxies binned by mass and SFR described in Table \ref{tab:mbin}, each pair of low and high SFR for a certain mass is connected by a black arrow. The objects identified via X-rays are shown with open squares (AGN) and diamonds (galaxies). The black dashed line represents the empirical division between star-forming galaxies (below the line) and galaxies with AGN (above the line) for galaxies at this redshift \citep{juneau2014}. Galaxies with lower stellar mass tend to have higher integrated $\OIIIHb$ ratio and also tend to have a higher inner  $\OIIIHb$ ratio. }
    \label{MEx}
\end{figure*}

We place the final sample of galaxies on a Mass-Excitation (MEx) diagram using the \cite{juneau2014} AGN/SF line shown in Figure~\ref{MEx}, with all the well-measured galaxies in the left panel and the galaxies with limits in their inner minus outer $\OIIIHb$ in the right panel. The median uncertainty in the integrated emission-line ratio is also shown in the lower left corner. Also shown are the objects in the sample that are X-ray sources and with X-ray emission classified as coming from AGN or star-forming galaxies \citep{luo2017,xue2016}. The triangles in the right panel represent upper or lower limits for the \OIIIHb\ \emph{gradient} as shown by the color bar. The black filled circles in the right hand panel are galaxies that have the same type of limit in both the inner and outer regions making the gradient unconstrained. The stars and hexagons are the ratios of the galaxies binned by mass and SFR described in Table \ref{tab:mbin}.

Figure \ref{MEx} shows that low-mass galaxies prefer higher $\OIIIHb$ in their nucleus. This will be further investigated in Section 5.2. The binned medians show no significant difference in the gradient between the high and low SFRs. 
The bins, similar to individual galaxies, indicate that low stellar mass galaxies marginally prefer nuclear ionization, while high stellar mass galaxies have neutral or slightly off-nuclear ionization. 

X-ray AGN are identified in massive galaxies, due to the well-known stellar mass bias for X-ray detection \citep{aird2012}. 
The most luminous known X-ray AGN do not have large \OIIIHb\ gradients, as the AGN in these galaxies are likely so bright that the emission from the AGN narrow line region overwhelms the entire host galaxy's star formation emission lines. These X-ray AGN are generally identified in traditional BPT diagrams because they have high integrated \OIIIHb\ ratios. This effect was also seen in the simulations of \cite{bridge2015}. 

\begin{figure*}[t]
\centering
\epsscale{0.9}
\plotone{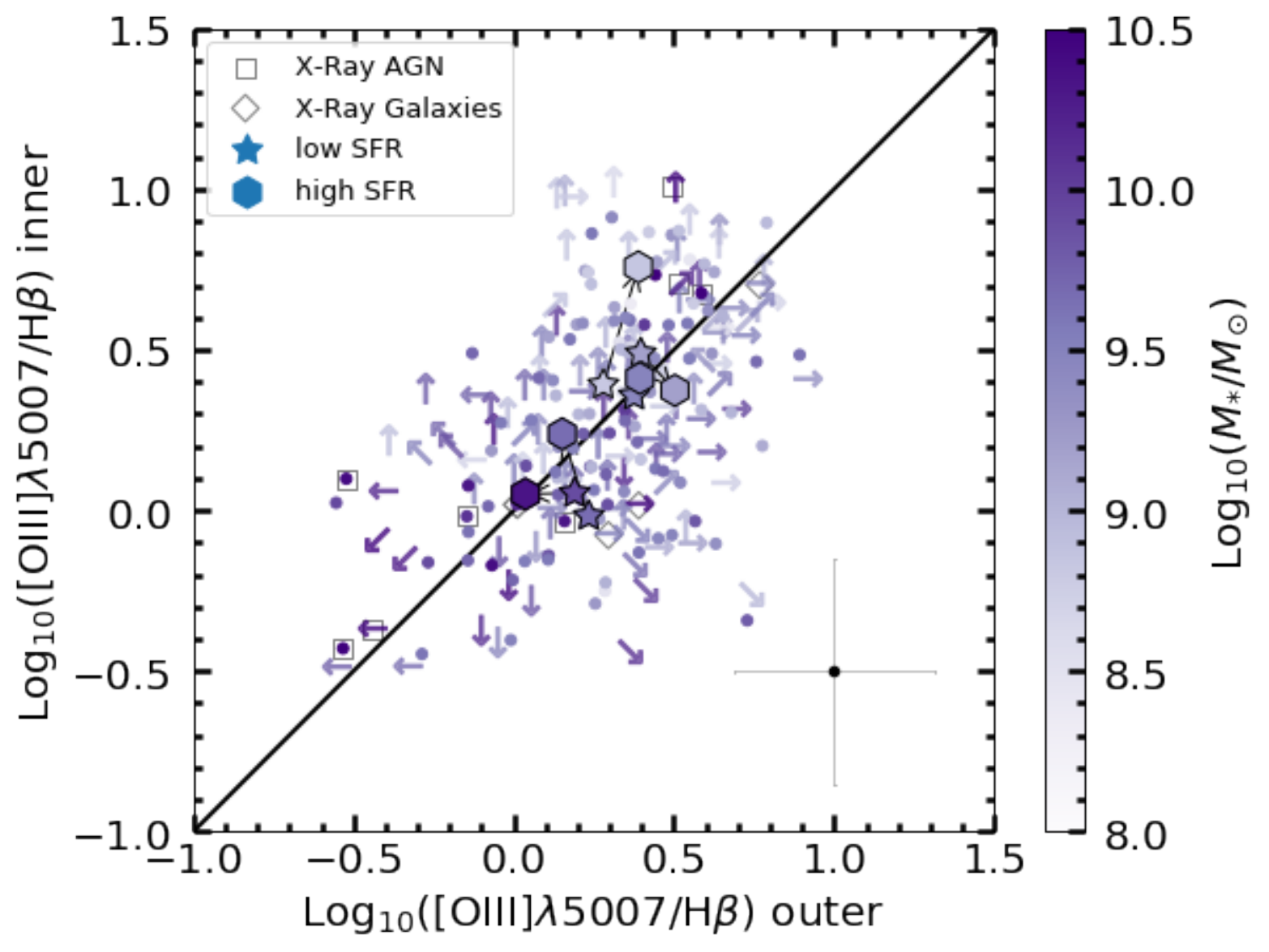}
\caption{Inner \OIIIHb\ vs outer \OIIIHb\ color-coded by stellar mass. The arrows represent limits in the measured line ratios, with diagonal arrows representing limits in both axes. The median uncertainty for the inner and outer ratios is shown in the lower right of the panel. The black line is the 1:1 ratio between the inner and outer regions. The stars and hexagons are from the bins described in Table \ref{tab:mbin}. All the bins are within 1$\sigma$ of having a zero gradient.}
\label{fig:invout}
\end{figure*}

Figure \ref{fig:invout} further investigates the relationship between \OIIIHb\ gradient and galaxy stellar mass. There is no strong trend between the resolved \OIIIHb\ and stellar mass, as there is a large range of \OIIIHb\ gradients across the sample. There is a marginal preference for low mass galaxies to have nuclear ionization. 
In sections 6.2 and 6.3 we further investigate the galaxies which deviate from zero gradient.

We now investigate how the spatially resolved emission-line gradient is affected by a galaxy's classification as a star-forming or AGN galaxy, as well as how the \OIIIHb\ gradient correlates to galaxy stellar mass, SFR, size, and redshift.

\begin{figure*}[t]
\centering
\epsscale{1.1}
\plotone{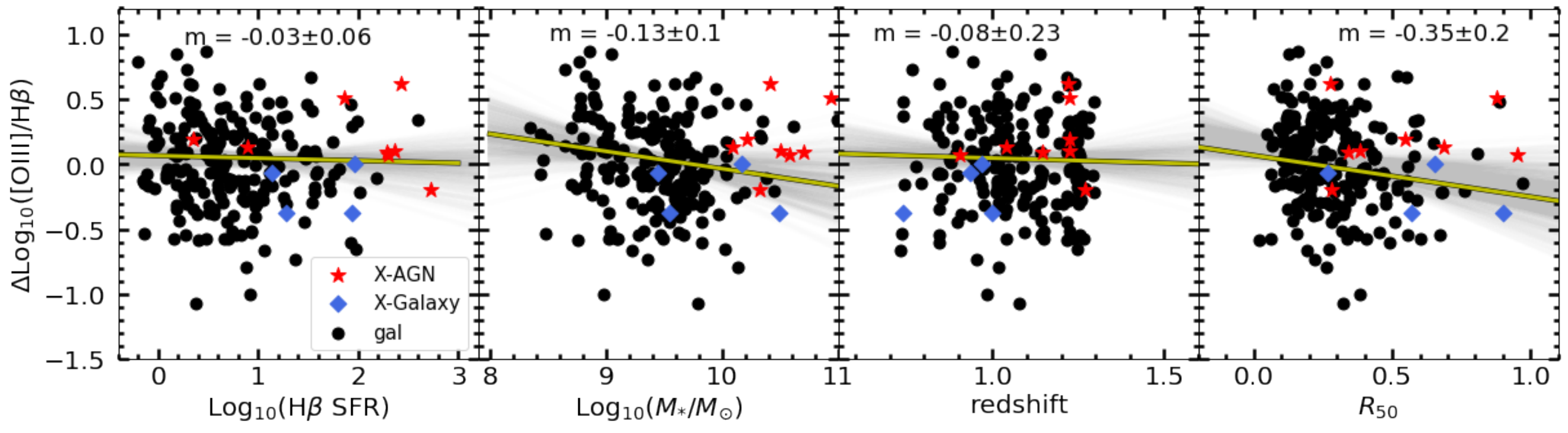}
\caption{The relationships of $\OIIIHb$ gradient with galaxy SFR, stellar mass, redshift, and effective radius. X-ray AGN and X-ray galaxies are shown as red stars and blue diamonds, respectively. Each panel has a yellow linear regression best-fit line, fit to the galaxies excluding X-ray AGN, with its slope showing the strength of the relationship. The $\OIIIHb$ gradient has marginal anti-correlations with stellar mass and effective radius.} 
\label{fig:galprop1}
\end{figure*}

Figure \ref{fig:galprop1} investigates the relationship between the line-ratio gradient and galaxy properties, with X-ray AGN and X-ray galaxies shown as red stars and blue diamonds respectively. We carry out a linear regression fit to the sample excluding the X-ray AGN and plotted the best-fit line in yellow with its slope and uncertainty reported in each panel. In all of the panels, X-ray AGN have a preference for higher $\OIIIHb$ in the nucleus, which we would expect for an AGN ionizing the central gas. The average \OIIIHb\ gradient of X-ray AGN is 0.12~dex, where the rest of the galaxy population has a  \OIIIHb\ gradient of -0.01~dex. X-ray galaxies, on the other hand, have similar \OIIIHb\ gradients to other galaxies of the same stellar mass.

In Figure \ref{fig:galprop1}, we find that the gradient does not have a relationship with $\Hb$ SFR or with redshift, as shown in the first and third panels. There is a marginal (1.55$\sigma$) anti-correlation between gradient and stellar mass, which was also observed in Figure \ref{MEx}. There is also a marginal (2.5$\sigma$) relationship between line-ratio gradient and galaxy size, with higher nuclear ionization in smaller galaxies. The relationships of line-ratio gradient with stellar mass and size are likely related due to the known correlation between galaxy mass and radius \citep{vanderwel2014,Yang2021,Mowla2019}.

\section{Discussion}

\subsection{An Intrinsic Population of Galaxies with Higher Nuclear Ionization}

Figure 3 shows a broad distribution of measured \OIIIHb\ gradients, and Figure 7 shows that nuclear ionization is marginally preferred in X-ray AGN, low-mass galaxies, and small galaxies. But it is unclear how much of the distribution of measured gradients is due to noise or due to the intrinsic galaxy population. To further investigate this, we compare our measurements with different models for the underlying distribution.

\begin{figure*}[t]
\centering
\epsscale{1.1}
\plotone{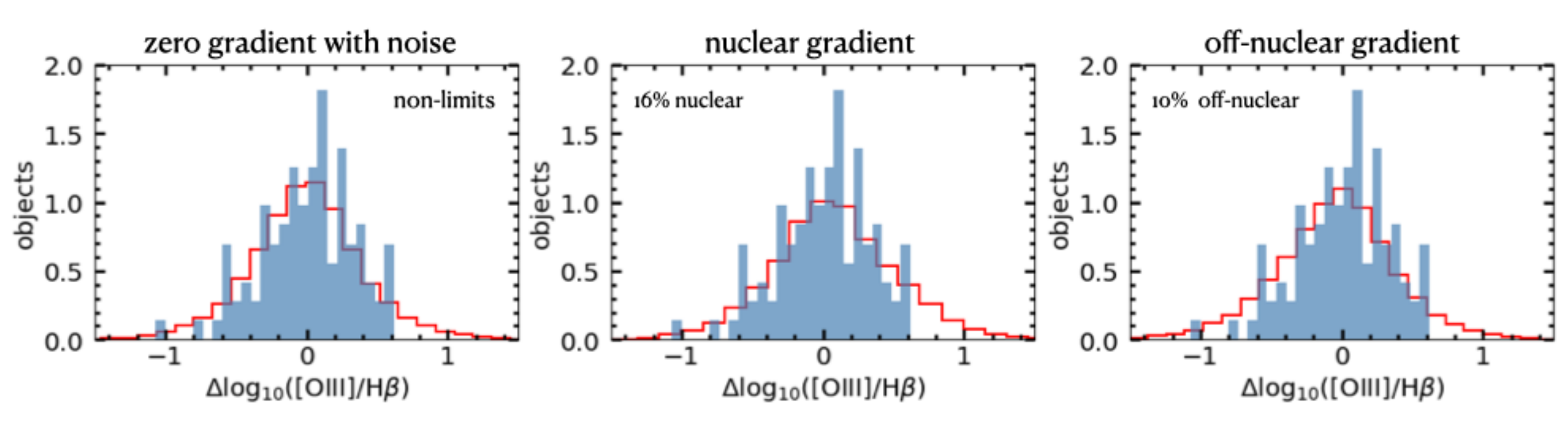}
\caption{
In each panel we compare the distribution of emission-line ratio gradients in our well-measured galaxies 
(blue histogram in each panel) to simulated datasets (red line). In the left panel we show a simulated distribution with zero emission line gradients with noise. An AD test returns $p>0.25$, 
showing the distributions are consistent. In the middle panel we increase the fraction of galaxies with higher nuclear emission-line gradients (+0.5~dex) until the AD-test is $p<0.05$: this occurs where 16\% of the galaxies have gradient (with 0.5~dex higher \OIIIHb\ in the nucleus). The right panel repeats the test but using -0.5~dex gradients, where we find that the AD-test is $p<0.05$ when 10\% of the galaxies show off-nuclear gradients (where the \OIIIHb\ ratio is 0.5~dex higher in the outskirts of the galaxies).}
\label{Fig:model1}
\end{figure*}

Figure \ref{Fig:model1} shows galaxies with well-measured line-ratio gradients as the blue histogram, and compares its distribution to different models shown as a red line in each panel. In the left panel our model depicts a distribution of galaxies with zero intrinsic gradient and with noise drawn from the observed uncertainties. An Anderson-Darling test returned a p-value of $>$0.25, 
indicating that the model is consistent with being drawn from the same parent distribution as the observed galaxies. In the middle and right panels we start adding galaxies with higher nuclear or off-nuclear \OIIIHb\ ratios (a gradient of +0.5~dex and -0.5~dex, respectively), until an Anderson-Darling test finds $p<0.05$ such that the distributions are no longer consistent with the same parent distribution. We choose $\pm 0.5$~dex because, as shown in Figure \ref{fig:grad_uncer}, our well-measured gradient sample (the solid blue histogram) uncertainty extends to 0.5~dex. The middle panel indicates that up to $16\%$ of the galaxies can have nuclear \OIIIHb\ that is 0.5~dex higher than the outer region before the model has a different parent distribution from the sub-sample. In the right panel, the model distributions indicate that up to 10\% of the galaxies can have off-nuclear ionization before it has a different distribution from the sub-sample. This shows that even though our sub-sample distribution is consistent with noise there is still room for a significant fraction of the galaxies to have large nuclear or off-nuclear ionization profiles.

\begin{figure*}[t]
\centering
\epsscale{0.8}
\plotone{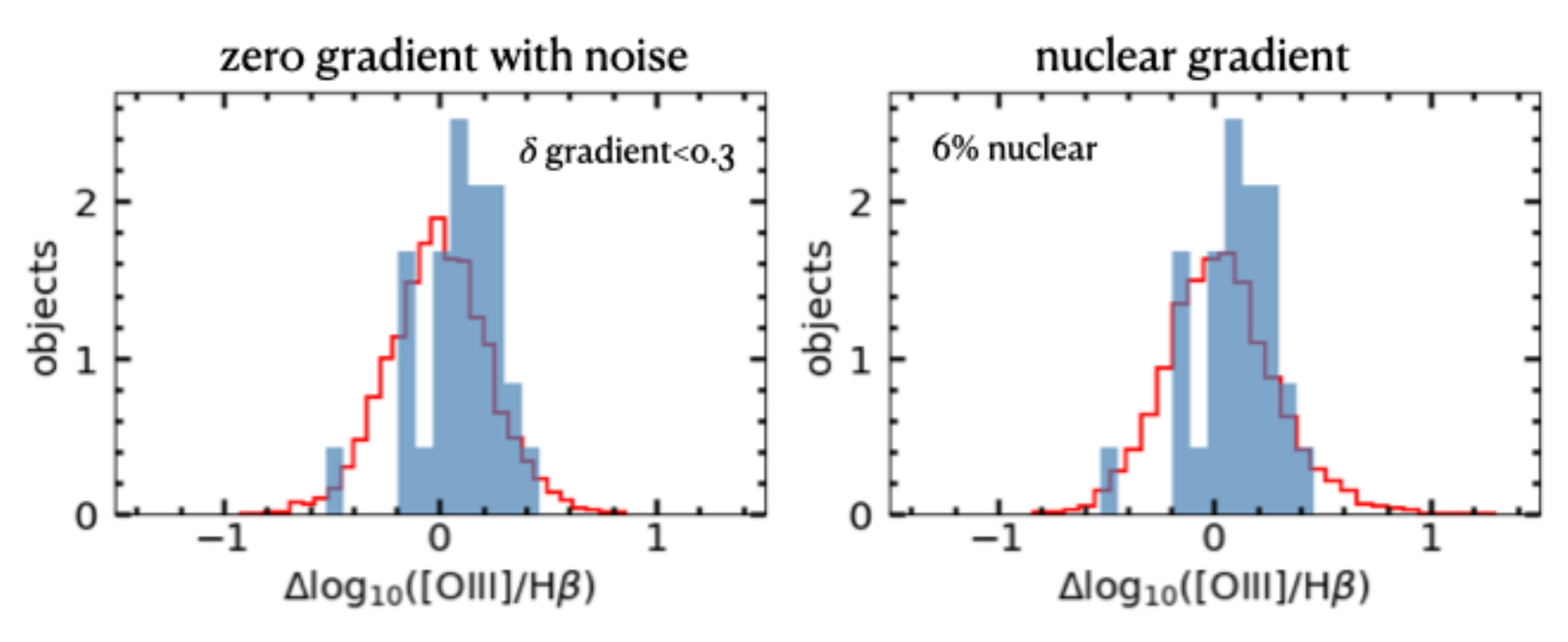}
\caption{
In each panel we compare the distribution of emission-line ratio gradients in our galaxies with a $\delta \Delta \log_{10}(\OIIIHb)$ under 0.3~dex (blue histograms in each panel) to simulated datasets (red line). In the left panel we show a simulated distribution with zero emission line gradients with noise. An AD test returns p=.01, showing the distributions are inconsistent. In the right panel we increase the fraction of galaxies with strong emission-line gradients (+0.5 dex) until the AD-test is p > 0.05:  this occurs where 5.5\% of the galaxies have gradient (with stronger OIII/H-beta in the nucleus). This tells us the the off-nuclear ionization galaxies are more likely to just be caused by noise.}
\label{Fig:model2}
\end{figure*}

We repeat this analysis again but this time with a subsample of galaxies that have the lowest $\delta \Delta \log_{10}(\OIIIHb)$, shown in Figure \ref{Fig:model2}.
This leaves us with 30 galaxies in our blue histograms with a gradient uncertainty less than 0.3~dex. The left panel shows a model with zero intrinsic gradient and the same uncertainties as the observations. In this case, an Anderson-Darling test indicates that model is not a good description of the data, with $p=0.01$. This indicates that, at least among galaxies with the smallest uncertainties in the line-ratio profiles, there is a significant population of galaxies with intrinsically higher nuclear \OIIIHb\ that cannot be simply explained by noise. The right panel shows that our sample is best modeled by at least $6\%$ of the galaxies having nuclear $\OIIIHb$ that is 0.5~dex higher than the outer region. 
The best-measured galaxies are statistically inconsistent with all galaxies having a flat \OIIIHb\ profile, and $6\%$ of galaxies with 0.5~dex higher nuclear \OIII\ need to be added to reproduce the observations. Meanwhile, in the larger population we find we can fit strong \OIIIHb\ nuclear ionization (0.5~dex) in up to $16\%$ of galaxies. There appears to be a genuine intrinsic population of galaxies with higher ionization in their central regions as $6-16\%$ of galaxies have nuclear ionization that is 0.5~dex higher than their outer regions. As we used a p-value of 0.05 this 6-16\% fraction of galaxies with higher nuclear ionization roughly corresponds to a $95\%$ confidence interval.

\subsection{Nuclear Ionization and Low-Luminosity AGN}

\begin{figure*}
    \centering
    \includegraphics[width=0.75\textwidth]{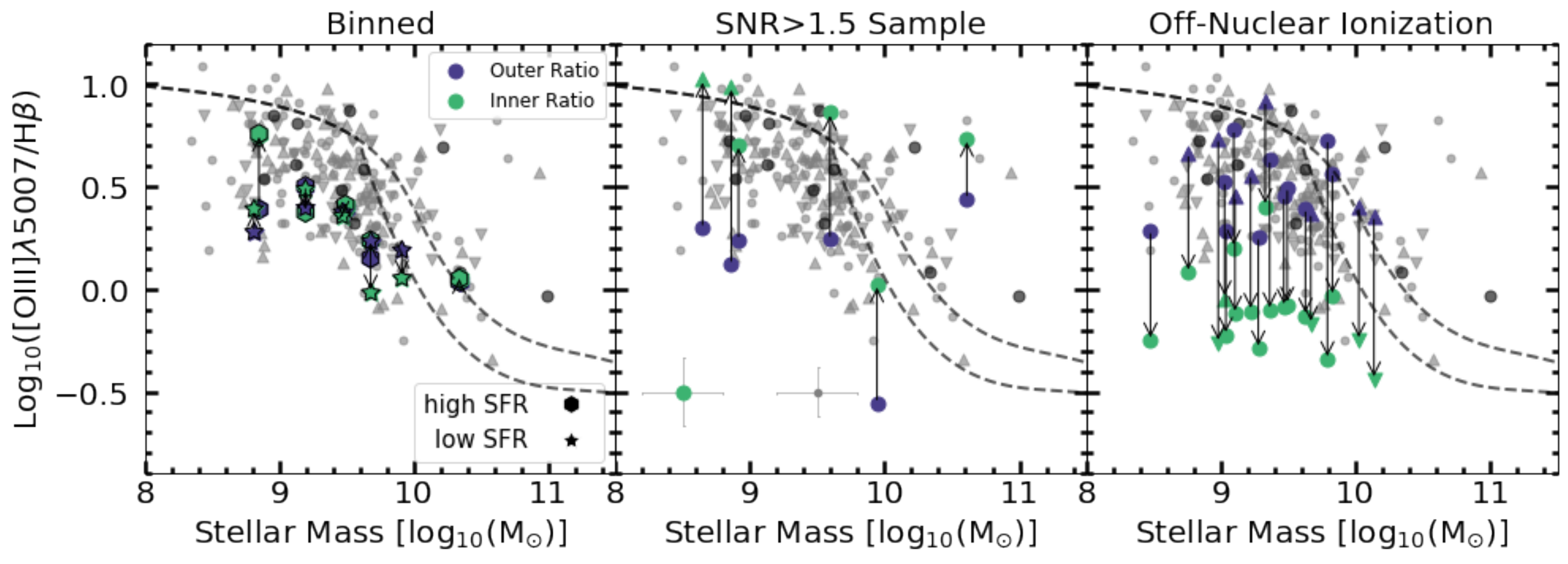}
    \caption{\emph{Left}: The inner and outer \OIIIHb\ ratios for galaxies binned by stellar mass and SFR as described by Table \ref{tab:mbin}. In all bins the inner and outer \OIIIHb\ ratios are consistent within $1\sigma$, suggesting that galaxies on average have relatively flat \OIIIHb\ gradients.
    \emph{Middle}: The six galaxies with nuclear \OIIIHb\ ratios that are at least 1.5$\sigma$ higher than their outer \OIIIHb\ ratios, suggesting higher nuclear ionization. All six of these galaxies are not X-ray detected in the deep CDF-S and CDF-N catalogs \citep{luo2017,xue2016}. \emph{Right}: 19 objects that have ($\sim$ 0.5 dex) higher \OIIIHb\ ratios in their outer regions compared to their inner ratio. }
    \label{fig:movers}
\end{figure*}

In the previous subsection we find evidence that 6-16\% of the galaxies in our sample have higher nuclear \OIIIHb. In this subsection we investigate individual galaxies that are candidates for having high nuclear ionization. The middle panel of Figure \ref{fig:movers} shows the line-ratio profiles of the 6 galaxies with positive \OIIIHb\ gradients detected with more than 1.5$\sigma$ significance. Two of these galaxies have a lower limit in their inner \OIIIHb\ ratio, as shown by the triangles in Figure \ref{fig:movers}. The galaxies shown in the middle panel of Figure \ref{fig:movers} have an average gradients of 0.6~dex, which is larger than the average X-ray AGN \OIIIHb\ gradient of 0.12~dex shown in Figure \ref{fig:galprop1} and the AGN identification threshold of $\Delta \log(\OIIIHb)= 0.1$~dex used in \cite{bridge2015}. This makes the six galaxies prime candidates to host low-luminosity AGN.

\begin{figure*}[t]
\centering
\epsscale{0.9}
\plotone{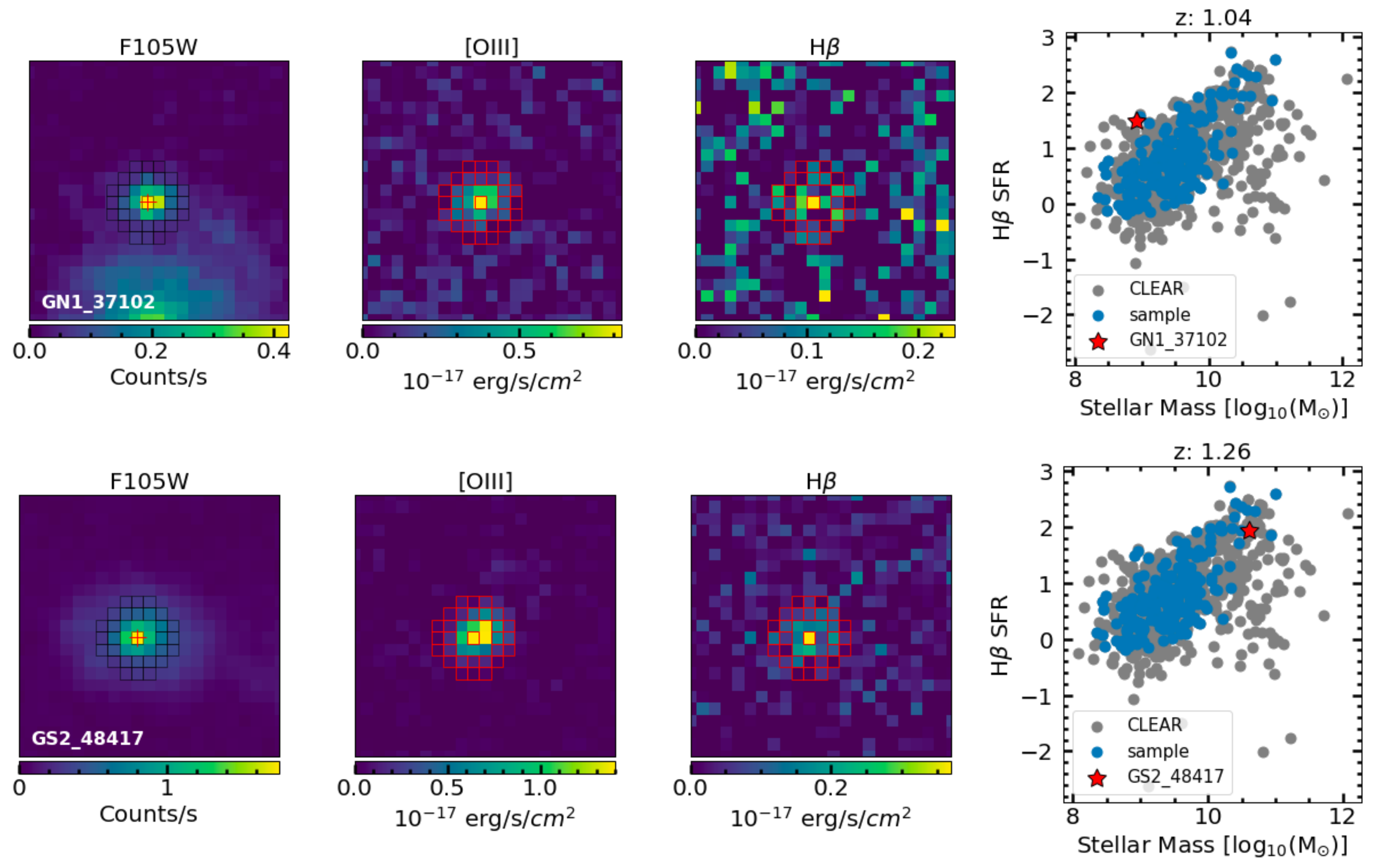}
\caption{Two of the galaxies from the right panel of Figure \ref{fig:movers} which have gradients with SNR>1.5 and indicate nuclear ionization. The left three panels show the F105W image, the \OIII\ emission-line map, and the \Hb\ emission-line map for each galaxy. \emph{Right}: The SFR and stellar mass of the parent sample with the galaxy of interest highlighted as a red star. Both galaxies appear to have $\OIII$ emission that is more compact than the continuum image and than the Hb emission.}
\label{Fig:AGN_angle}
\end{figure*}

Figure \ref{Fig:AGN_angle} shows continuum (F105W), $\OIII$, and $\Hb$ images for two of the galaxies identified in the right panel of Figure \ref{fig:movers}. The right panel marks the galaxy of interest in the distribution of SFR and stellar mass of the parent sample.  Figure \ref{Fig:AGN_angle} gives an example of a low-mass and a high-mass galaxy, where both are extended in their continuum and $\Hb$ images but have more compact $\OIII$ emission. The detection of higher nuclear \OIIIHb\ emission-line ratios could indicate a low-luminosity and/or obscured AGN that is undetected by the deep X-ray observations \citep{xue2016,luo2017}.

Gradients in \OIIIHb\ can also be caused by non-flat metallicity gradients resulting in varying \OIIIHb\ ratios across the disks. \cite{simons21} showed that high stellar mass galaxies at redshifts $z\sim1.5$ generally have flat metallicity profiles while low stellar mass galaxies have lower metallicity in the center (higher [OIII]/Hb in the center at the ~0.1 dex/kpc scale). This would cause higher $\OIIIHb$ in the center, however the gradients we measure are $\sim$0.5~dex over $\sim$2~kpc while \cite{simons21} have gradients $\sim$0.1 dex/kpc, so metallicity alone is unlikely to fully account for the elevated central $\OIIIHb$.


Using a similar analysis with stacked data from both CANDELS WFC3 and \textit{Chandra}, \cite{trump2011} found that it was likely that at least some of the galaxies in their sample of 28 objects contained weak AGN. Single-object studies have also been done: \cite{wright2010} used the OSIRIS integral field spectrograph on the Keck Observatory to show that the nuclear region ($\sim0$\farcs2) of a $z\sim1.5$ galaxy exhibits elevated \OIIIHb\ and [N~II]/H$\alpha$ ratios, which they posit is an indication of an embedded AGN. Our analysis finds 6 galaxies at $z\sim1$ which show promise for hosting low-luminosity AGN.  

\subsection{Off-Nuclear Ionization}

Figure \ref{MEx} and \ref{fig:invout} show a wide range of line-ratio profiles, including galaxies with higher $\OIIIHb$ in their outer regions. Section 6.1 indicates that the observed high off-nuclear \OIIIHb\ ratios are consistent with flat \OIIIHb\ profiles and observational uncertainties. But the uncertainties are sufficiently large that the same analysis indicates that up to 10\% of the galaxies might have off-nuclear ionization profiles with \OIIIHb\ that is 0.5~dex higher in galaxy outskirts than in galaxy centers. Such systems could be off-nuclear AGN. Evidence of off-nuclear massive black holes has been observed near the center of the Milky Way \citep{oka2017}, and the discovery of extragalactic off-nuclear X-ray sources \citep[e.g.,][]{farrell2009, jonker2010, barrows2016} has lent credence to the idea that massive black holes exist at significant, 1~kpc or greater, distances from the centers of galaxies.

\begin{figure*}[b]
\centering
\epsscale{0.9}
\plotone{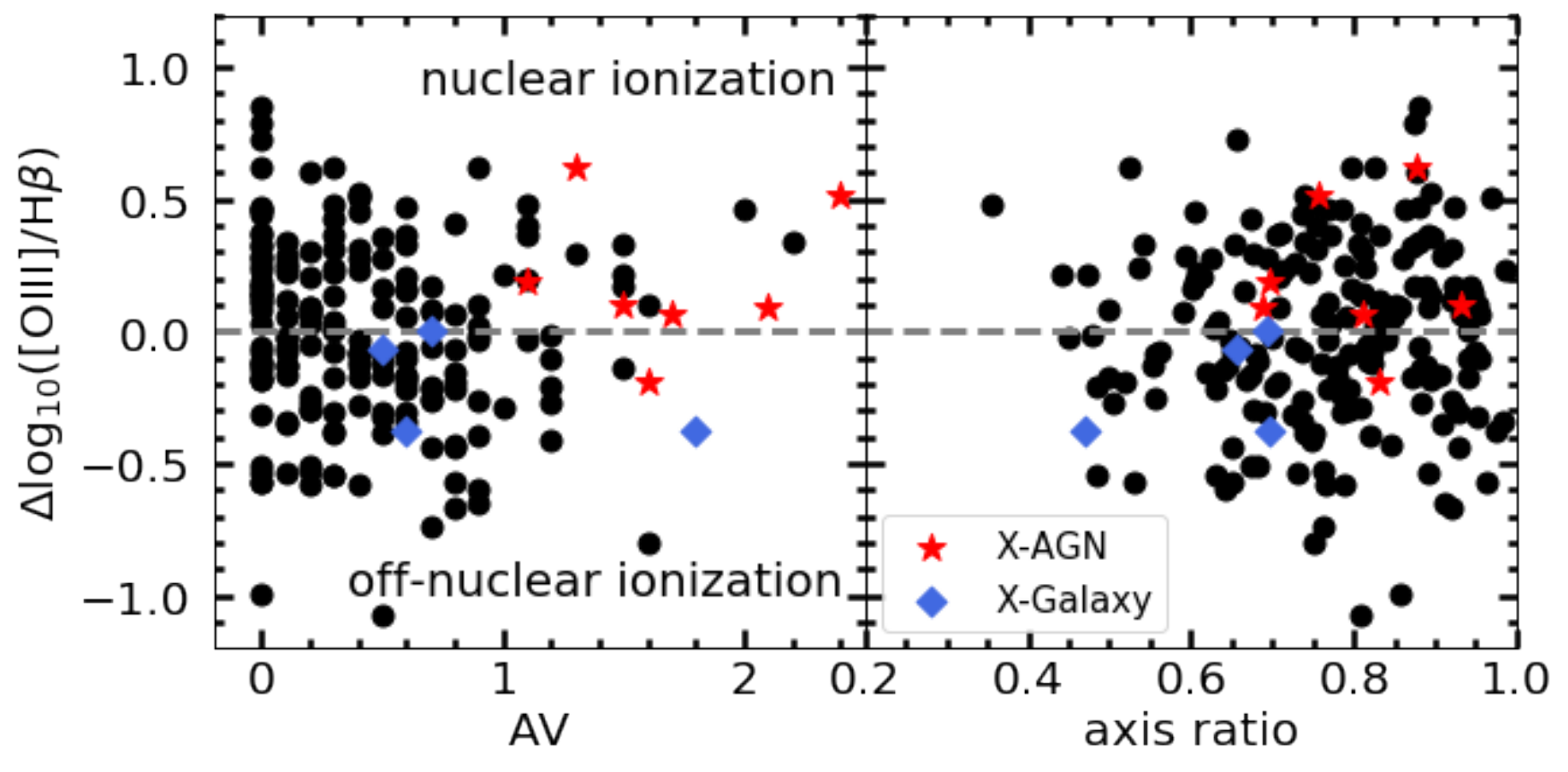}
\caption{\textit{Left:} The gradient of inner minus outer $\OIIIHb$ versus dust attenuation. \textit{Right:} The gradient of $\OIIIHb$ versus the galaxy axis ratio. X-ray AGN and X-ray galaxies are red stars and blue triangles, respectively. The off-nuclear ionization galaxies (negative $\Delta \log(\OIIIHb)$) do not have more dust attenuation that the rest of the sample, and also are not preferentially edge-on galaxies. An Anderson-Darling test between the off-nuclear sources and the full sample returns a p-value $>0.25$, indicating the high off nuclear ionization galaxies are consistent with the same parent distribution as the rest of the sample.}
\label{Fig:dust}
\end{figure*} 

\begin{figure*}[t]
\centering
\epsscale{1.0}
\plotone{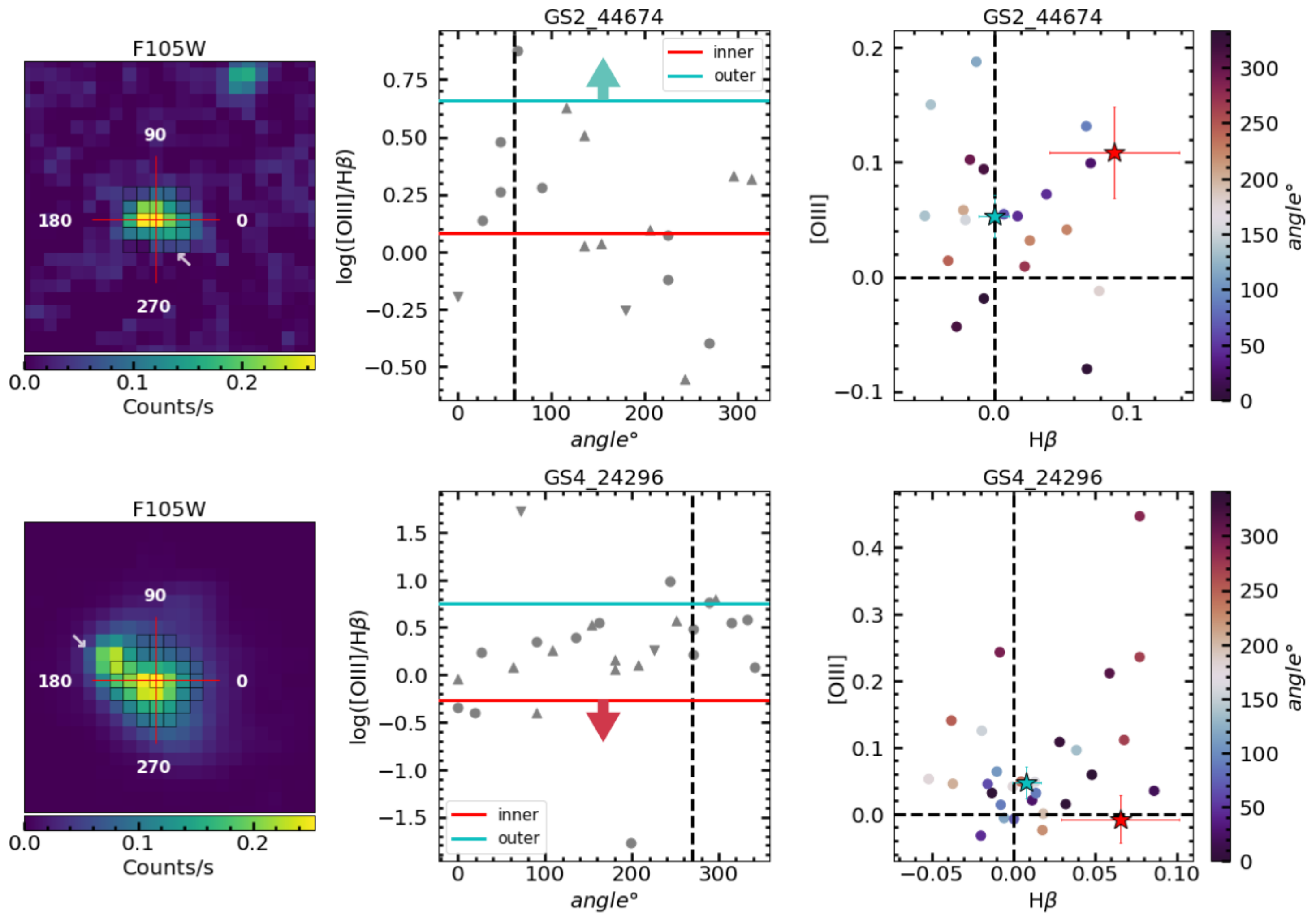}
\caption{Two galaxies with higher off-nuclear \OIIIHb\ in the outskirts than in their centers. The left panels show the direct F105W image. The center panels show the \OIIIHb\ of the outer-region pixels as a function of the angle shown in the left panel, with the inner-pixel ratio and median outer-pixel ratio shown as horizontal lines. The right panels show the outer-pixel \OIII\ and \Hb\ fluxes color-coded by angle, with limits in line fluxes and ratios shown by triangles. The inner-pixel ratio and median outer-pixel ratio are shown as red and blue stars, respectively. Both galaxies have an apparent preference for high \OIIIHb\ ratios at particular angles, shown by dashed vertical lines in the middle panel, 
that do not coincide with the continuum light in the direct image, indicated by white arrows in the left panel. }
\label{Fig:angle}
\end{figure*} 

We investigate the possibility of off-nuclear AGN by investigating the line-ratio maps of a sample of 19 objects that have ($\sim$0.5~dex) higher \OIIIHb\ ratios in their outer regions compared to their inner ratio as shown in the right panel of Figure \ref{fig:movers}. Of the 19 galaxies, five of them cross the AGN/SF line with an outer region indicating an AGN while the central ratio stays in the SF region. Three of these five points also have a lower limit in their outer ratio measurement.

We first investigate if dust and/or orientation plays a role in causing higher off-nuclear \OIIIHb\ ($\Delta \OIIIHb$<0), using measurements of $A_V$ attenuation and axis ratio from the compilation of \cite{barro2019}. Matching catalogs results in 196 galaxies from our sample, for which we show AV and axis ratio measurements in Figure \ref{Fig:dust}. From the left panel we can see that the off-nuclear ionized galaxies do not have more dust attenuation that the rest of the sample. The right panel shows that the off-nuclear ionized galaxies are not preferentially viewed edge-on. An Anderson-Darling test between the off-nuclear sources and the full sample returns a p-value $>0.25$ for both dust attenuation and axis ratio, showing the high off-nuclear ionization galaxies have the same distribution as the rest of the sample. This shows that high \OIIIHb\ ratios in galaxy outskirts are not preferentially produced by dust attenuation.

We investigate the possibility of these galaxies hosting off-nuclear AGN by looking for clumping of off-nuclear \OIIIHb\ and/or continuum light at particular angles, as shown in Figure \ref{Fig:angle}. The left panel shows the F105W continuum image, the middle panel shows the emission-line ratio of each pixel in the outer region verses the angle, and the right panel shows the emission-line flux of each outer-region pixel color-coded by the angle. The middle and right panel show the median for the inner (outer) regions as a red (magenta) line or star, respectively, and limits are shown by triangles. Off-nuclear AGN might lead to a clump of high \OIIIHb\ ratio and high \OIII\ flux at a narrow range of angle. This is shown by simulations done by \cite{bellovary2010} and \cite{See22} which indicate that off-nuclear AGN can be detected if they retain a bound clump of both gas and stars.

These are the only 2 of the 19 galaxies with high off-nuclear $\OIIIHb$ that have a preferred angle for higher \OIIIHb\ emission. If we also see more continuum light at that angle, this would give further indication of an off-nuclear AGN. However, we did not see this for either galaxy. The two galaxies shown in Figure \ref{Fig:angle} do show a peak $\OIIIHb$ ratio at a particular angle in the middle figure, at $\sim 60 \deg$ for the top panel and $\sim 260 \deg$ in the bottom panels. But the continuum light in these galaxies does not have the same distribution, and is instead concentrated at different angles (330 deg in the top galaxy, and 150 deg in the bottom). The other 17 galaxies with high outer-region \OIIIHb\ do not show this kind of clumpiness, but instead the \OIIIHb\ is distributed over a wide range of angles. 

The two galaxies examined in Figure \ref{Fig:angle} do not seem to be well-described as off-nuclear AGN. These example galaxies could instead be ionized outflows, as there is not a concentration of continuum light in the areas that are marked with high ionization. High outer-region \OIIIHb\ could indicate a metal-rich nucleus and metal-poor disk (commonly referred to as a "negative metallicity gradient"), as observed in some massive high-redshift galaxies \citep{Wuyts2016}. The number of galaxies with off-nuclear ionization is also fully consistent with the large uncertainty in the measured gradients, as indicated from our models of the intrinsic distribution of line-ratio gradients.

\section{Conclusions}

We report measurements of resolved \OIIIHb\ line-ratio profiles via \emph{HST}/WFC3 G102 grism observations taken as part of the CLEAR survey. We measured \OIIIHb\ gradients by taking the center of the galaxy, defined by the photometric center of the F105W image, and subtracting the median of the \OIIIHb\ ratio in an annular extended region. We investigate the spatially-resolved nature of this line ratio, and we summarize our results as follows:

\begin{itemize}

\item The data binned by stellar mass and SFR in Figures \ref{MEx} and \ref{fig:invout} show that low stellar mass galaxies tend to have marginally higher nuclear $\OIIIHb$ whereas high stellar mass galaxies have flat gradients. The SFR of the galaxy is not correlated with the $\OIIIHb$ gradient. A slight anti-correlation ($\sim 1.6\sigma$) between stellar mass and $\OIIIHb$ gradient is shown in Figure \ref{fig:galprop1}.
This also shows a $\sim2\sigma$ anti-correlation of effective radius with $\OIIIHb$ gradient.

\item X-ray AGN have a 0.1~dex higher nuclear $\OIIIHb$ than the rest of the sample, as seen in Figure \ref{fig:galprop1}.

\item We compare toy models to the distribution of galaxies with well-measured gradients and galaxies with the lowest gradient uncertainty from our data set. From these toy models, Figures \ref{Fig:model1} and \ref{Fig:model2}, we found that 
6-16\% of the sample is likely to include galaxies with higher nuclear \OIIIHb. The lower limit comes from the number of galaxies needed for the model to have a similar distribution to galaxies with the smallest gradient uncertainty, while the upper limit is from the number of galaxies before the model has a different distribution from galaxies with no limit in its gradient measurement.

\item The toy models show that galaxies with higher nuclear ionization are likely to exist, but the individual detections are marginal (only $\sim$1.5$\sigma$) as shown in Figure \ref{fig:movers}. 
These galaxies potentially host low-luminosity AGN in its nucleus that are not detectable by X-rays or BPT diagrams.
 
\item We investigate the galaxies with the largest measured off-nuclear \OIIIHb\ ratios and find that they are mostly consistent with observational noise. Two galaxies shown in Figure \ref{Fig:angle} have "clumpy" off-nuclear \OIII\ emission that is roughly orthogonal to the morphology of the continuum image, and so is probably better described by an ionized gas cone than by off-nuclear AGN.

\end{itemize}

This work sheds light on the ensemble properties of 1<z<2 galaxies, especially the conclusion that 6-16\% of galaxies have considerable ($\sim$0.5~dex) central \OIIIHb\ gradients indicative of nuclear AGN that are missed by other methods. But the measurements of individual galaxies are limited by large uncertainties in the CLEAR data. A better understanding of individual galaxies, including searches for off-nuclear AGN or more nuanced ionization profiles, requires deeper data. This is available from HST/WFC3 grism data in FIGS and MUDF, and will soon be available from JWST/NIRISS surveys like NGDEEP.

\section{Acknowledgments}

This work is based on data obtained from the Hubble Space Telescope through program number GO-14227. Support for program GO-14227 was provided by NASA through a grant from the Space Telescope Science Institute, which is operated by the Association of Universities for Research in Astronomy, Incorporated, under NASA contract NAS5-26555. B.E.B.\ and J.R.T.\ acknowledge support from NSF grant CAREER-1945546 and NASA grant JWST-ERS-01345. J.S.B.\ acknowledges support from NASA/STScI grant HST-AR-15008.

\facilities{\textit{HST} (WFC3)}

\software{\texttt{AstroPy} \citep{astropy2013}, \texttt{Matplotlib} \citep{hunter2007}, \texttt{NumPy} \citep{vanderwalt2011}, \texttt{SciPy} \citep{jones2001}, \texttt{grizli} \citep{grizli}, \texttt{eazy-py} \citep{brammer2008, eazy-py}, \texttt{FSPS} \citep{conroy2010}}

\clearpage

\bibliography{spatial_grism}

\end{document}